\documentclass[oneside,letterpaper]{amsart}
\usepackage[latin1]{inputenc}
\usepackage{amssymb}

\usepackage[nobysame]{amsrefs}
\usepackage{geometry}
\usepackage{float}
\usepackage{graphicx}

\usepackage{appendix}

\makeatletter
\theoremstyle{plain}
\newtheorem{thm}{Theorem}[section]
\theoremstyle{plain}

\theoremstyle{remark}
\newtheorem*{acknowledgement*}{Acknowledgement}
\theoremstyle{definition}
\newtheorem*{defn*}{Definition}
\newtheorem{defn}[thm]{Definition}

\makeatother
\begin{document}

\title{Dual Non-Abelian Yang-Mills Simulations in Four Dimensions}
\author{J. Wade Cherrington$^1$}
\maketitle

\vspace*{-10pt}
\begin{center}
\small
$^1$Department of Mathematics, University of Western 
Ontario, London, Ontario, Canada \newline
\end{center}

\begin{abstract}
We present numerical results for pure $SU(2)$ Yang-Mills theory in four space-time 
dimensions using a novel algorithm based on dually transformed variables. The simulation 
makes use of a recently derived $O(j^4)$ algorithm for the dual vertex amplitude and a dual Metropolis
algorithm that generalizes the one recently developed for three dimensions.
The dual algorithm is validated against the equivalent model using conventional 
variables over a range of couplings, spin cut-offs, and lattice sizes.
We consider a lattice size up to $8^4$, where the problem of negative amplitudes renders the simulation results
excessively noisy even at a relatively low $\beta$ (starting at about $\beta=1.8$).
In conclusion, we survey some approaches to addressing the sign problem and increasing the efficiency 
of dual computations.
\end{abstract}

\section{Introduction}
Exact duality transformations for statistical mechanical systems with many degrees of freedom
have provided a rich source of insights and alternative computational methods over the years~\cites{KW1941,CS2004}.
As lattice regularized quantum theories share much of the essential
structure of lattice statistical mechanical models, it is natural to investigate analogous duality transformations.
Indeed, applications of duality to quantum fields on the lattice continues to grow; recent examples 
include~\cites{ShCh08,Wolff09,W08II, W08I}.  In the present work, we shall focus on pure Yang-Mills theory 
as we expect it will be instructive to construct dual algorithms in this case before proceeding to 
dynamical fermions in the dual.

For lattice gauge theory (LGT) with Abelian groups such as $U(1)$ and $\mathbb{Z}_2$, it is well 
understood how to construct dual models useful for practical computations~\cite{Panero2004, Panero2005, PollyWiese, Savit80}.
However, dual simulations with non-abelian gauge groups have presented a greater challenge. The main difficulties 
have been the definition of the dual model in terms of practically computable amplitudes and the construction of 
an ergodic set of moves that can be used with a Markov chain Monte Carlo method (e.g. the Metropolis algorithm).

Following some initial results on amplitudes for non-abelian lattice models in the specific case of $D=3$
and $G=SU(2)$ in~\cite{AS,ACSM}, a procedure for obtaining non-abelian dual models in four
and higher dimensions on the lattice was given by Halliday and Suranyi~\cite{Halliday95}.
More recently, a new mathematical framework for understanding the non-abelian dual transformation was constructed,
the lattice spin foam formulation~\cite{OecklDGT, OecklPfeiffer}. With regard to numerical simulations, the spin foam approach
is helpful for computing amplitudes (particularly in the four-dimensional case~\cite{CC2009}) as it gives an explicit prescription
for the dual amplitudes in terms of spin networks. These spin networks can in turn be reduced in complexity
from their initial definition by the use of diagrammatic recoupling methods~\cite{KauffmanLins,CarterSaito}, thus 
avoiding the typically explosive proliferation of tensorial algebra that would otherwise ensue.  The spin foam
construction also makes more transparent the distinction between irrep labels associated to plaquettes and
intertwiner labels that are naturally associated to edges, as well as the admissibility constraints between 
these quantities necessary for a non-zero amplitude. In~\cite{CCK}, these features were applied to the three 
dimensional case with $G=SU(2)$, resulting in an explicit algorithm that was validated against conventional results.
Further details on the development and current status of this area can be found in~\cite{LAT08}.

The present work may be viewed as a follow-up in four dimensions to the previous results in three dimensions that 
includes~\cite{CCK} as well as earlier work by Hari Dass~\emph{et al.}~\cite{Dass83,Dass94,DassShin}. In 
the three-dimensional case, an explicit amplitude was known from~\cite{AS,ACSM} and the primary 
challenge was finding an ergodic set of moves. In the present case, the four-dimensional (hyper-cubic) vertex 
amplitude is a non-trivial function of 48 spin variables, and it wasn't until recently that a practical 
algorithm for this vertex amplitude~\cite{CC2009} was available.

In this paper, we construct a set of ergodic moves (detailed in the Appendix) which when combined with the amplitude of~\cite{CC2009}
allow us to perform the first Metropolis algorithm simulations of dual $SU(2)$ in four dimensions.
We emphasize that the primary purpose of the present work was to validate the dual algorithm and its implementation
against a conventional algorithm, in order to understand its performance over a range of couplings and to evaluate 
the severity of the sign problem in a simple setting. Efforts to extend the methods to more observables 
such as Wilson or Polyakov loops are presently underway by the author.

This paper is organized as follows. In Section 2, we briefly summarize the form of the dual 
model for an $SU(2)$ lattice gauge theory and define the observable to be computed. 
In Section 3, we describe how a conventional lattice code was constructed and used to 
compute expectation values of an appropriately chosen effective observable.  Section 4
summarizes the results obtained with the dual algorithm. The paper is concluded with some discussion
of the sign problem and other features of the dual algorithm in Section 5. Details of the moves
that define the dual algorithm are provided in the Appendix.

\section{The Spin Foam Simulations}\label{sec:spinfoamsims}
In the present section we shall briefly review some essential features of the dual spin foam model in order to describe 
the simulations performed and results obtained.  The reader is referred to~\cite{CCK} for a derivation of spin 
foam duality motivated by simulations with $G=SU(2)$ in three dimensions; its adaption to the present case of four dimensions
can be found in~\cite{CC2009}. A general derivation of spin foam models can be found in~\cite{OecklPfeiffer}.

In a dual spin foam model, the continuous group-valued variables of the conventional theory are replaced by 
discrete labels. The plaquettes of the lattice are labelled by irreducible unitary representations of $SU(2)$, 
typically taken as the half-integers (although whole integers are sometimes used in diagrammatic methods). 
For a given plaquette labelling, each edge carries an intertwiner label, 
corresponding to one of a complete basis of maps that intertwine the irreps on plaquettes that intersect the edge. If an
appropriate set of conditions is satisfied, than a particular labelling of plaquettes by irreps
and edges by intertwiner labels is said to form an \emph{admissible} spin foam. Further details on these
variables and the admissibility constraints among them are given in the Appendix.

As our present motivation is to evaluate the correctness and efficiency of the dual algorithm relative to the 
conventional formulation, we chose an observable that is particularly simple to compute in the 
dual model.  Such an observable is provided by the average value of the spin at a plaquette $p^{*}$:
\begin{equation}
\left< j_{p^{*}} \right>_{D} \equiv \frac{ \sum_{f \in F_{0}} j_{p^{*}}(f) \: \mathcal{A}(f) } { \sum_{f \in F_{0}} \mathcal{A}(f) },
\end{equation}
where $f \in F_{0}$ are the admissible spin foams, and $A(f)$ is the dual amplitude, as reviewed in
Appendix~\ref{sec:alg_append}. The observable $j_{p}$ is the spin label assigned by spin foam $f$ to plaquette $p$. 
The subscript $D$ on the LHS indicates evaluation of the observable against the dual (as opposed to conventional)
amplitude $A(f)$. In Section \ref{sec:Conventional}, we will describe how to define and compute the same observable within the
conventional LGT framework.

For concreteness, we define here precisely which amplitudes were used to generate the results given in Section~\ref{sec:results}. 
In terms of the spin foam amplitude $A(f)$ defined in equation~(\ref{eqn:partfun}) of the Appendix, the vertex
and edge amplitudes $A_{v}$ and $A_{e}$ are those defined in~\cite{CC2009}. The plaquette amplitude $A_{p}$
used comes from the \emph{heat kernel} action, defined as follows:

\begin{defn}[Heat Kernel Action]\label{defn:heatkernel}
Let $G=SU(N)$ and $g \in G$. Let $f(g,t)$ be the solution to the differential equation
\begin{equation}
\Delta f(g, t) = \frac {\partial f(g,t)} {\partial t}
\end{equation}
at $t=\frac{N}{\beta}$, with initial condition $f(g,0)=\delta(g)$. Then the action $S(g)$ defined implicitly by 
\begin{equation}
e^{-S(g)} = \frac{ f(g,t) } { f(I,t) }
\end{equation}
is known as the heat kernel action~\cite{Menotti}.
\end{defn}
It can be shown that the character expansion of the heat kernel action for $SU(2)$ is of the form
\begin{equation}\label{eqn:hk_expansion}
e^{-\beta S(g)} = \sum_{j} (2j+1)e^{-\frac{C_{j}}{2 \beta}} \chi_{j}(g)
\end{equation}
where $C_{j}$ is the quadratic Casimir eigenvalue associated to the $j$th representation; for
$SU(2)$, $C_{j} = j(j+1)$.

As the $\chi(g)$ are absorbed into the vertex and edge amplitudes of the dual model, by inspection
of~(\ref{eqn:hk_expansion}) the plaquette amplitude $A_{p}$ in equation~(\ref{eqn:partfun}) associated
with the heat kernel action is thus $A_{p}(j_{p})=(2j_{p}+1)e^{-\frac{C_{j_{p}}}{2 \beta}}$;
the reader is referred to~\cite{CCK} for further details.

\section{The Conventional Simulation}\label{sec:Conventional}
As we are working with a mathematically exact transformation between the conventional and dual
forms of the lattice gauge theory, it is possible to validate the correctness of the dual algorithm
by computing the same expectation value on both sides of the duality. We believe it is important 
to do so for several reasons. The dual vertex amplitude we make use of is relatively complex in both its 
derivation and final form~\cite{CC2009}. Assuming that a dual amplitude has been correctly derived and implemented, one would 
also like to verify that the algorithm constructed (see the Appendix) is in fact ergodic and capable of realizing convergence of
expectation values with comparable time and resources to that of a conventional code.  Agreement
within statistical error of conventional and dual results, while not constituting proof of a correct dual
algorithm, strongly indicates its correctness. Such testing is particularly valuable as the dual code is further optimized.

Given an observable and its expectation value defined in terms of the dual model, one can derive
the effective observable whose expectation value is equal when evaluated with respect to the 
original conventional model. Explicitly, we seek a function $O_{j_{p}}(g_{e})$ such that
\begin{eqnarray}\label{eqn:effectdef}
   \left< j_{p^{*}} \right>_{D} &=& \left< O_{j_{p^{*}}}\right>_{C} = \frac{1}{Z}
                                    \int \prod_{e \in E} dg_{e} \: O_{j_{p^{*}}}(g_{e}) e^{\sum_{p \in P}S(g_p)} \\ \nonumber 
                                &=& \frac{1}{Z} \int \prod_{e \in E} dg_{e} O_{j_{p^{*}}} \left( \prod_{p \in P}  \sum_{j_{p} \in \mathbb{Z^{+}}} 
                                     \: \text{dim}(j_{p}) c_{j_{p}} \chi_{j_{p}}(g_{p}) \right),
\end{eqnarray}
where we have introduced a character expansion in the second line to better show the relation with the dual observable.
We next recall from Section~\ref{sec:spinfoamsims} the dual observable of interest
\begin{eqnarray}\label{eqn:charsumj}
 \left< j_{p^{*}} \right>_{D}  &=&  \frac{1}{Z} \sum_{f \in F_{0}}  j_{p^{*}} \prod_{v \in V} A_{v}(\{j\}_v,\{i\}_v)
                           \prod_{e \in E} A_{e}(\{j\}_{e},i_e) \prod_{p \in P} A_{p}(j_{p}) \\  \nonumber
      &=& \frac{1}{Z} \int \prod_{e \in E} dg_{e} \left( \prod_{p \in P, p \neq p^{*}}  \sum_{j_{p} \in \mathbb{Z^{+}}} 
           \: \text{dim}(j_{p}) c_{j_{p}} \chi_{j_{p}}(g_{p}) \right) \sum_{j_{p^{*}} \in \mathbb{Z^{+}}} j_{p^{*}} 
           \: \text{dim}(j_{p^{*}}) c_{j_{p^{*}}} \chi_{j_{p^{*}}}(g_{p^{*}})
\end{eqnarray}
Based on the above we introduce the following proposal for $ O_{j_{p}}(g_{e})$:
\begin{eqnarray}\label{eqn:effectformula}
    O_{j_{p^{*}}}(g_{e}) &=& \left( \frac{\sum_{j_{p^{*}} \in \mathbb{Z^{+}}} j_{p^{*}} \: \text{dim}(j_{p^{*}}) c_{j} \chi_{j_{p^{*}}}(g_{p^{*}})}
                          {\sum_{j_{p^{*}} \in \mathbb{Z^{+}}} \: \text{dim}(j_{p^{*}}) c_{j} \chi_{j}(g_{p^{*}}) } \right).
\end{eqnarray}
The above can be checked by substituting into the second line of equation (\ref{eqn:effectdef}) and
comparing to the second line of equation (\ref{eqn:charsumj}).

Given the above, it is useful to be able to readily compute the character expansion components $c_{j}$ of the action in order to compute the
``re-weighted'' numerator of~(\ref{eqn:effectformula}). The heat kernel action used in the present simulations
has a simple form (see Definition~\ref{defn:heatkernel}) in which character coefficients can be rapidly evaluated. 
We note in passing that the division by part of the original
amplitude as is done in (\ref{eqn:effectformula}) in general may produce numerical problems, as convergence is better
when observables are well correlated with the amplitude. In practice, for this particular case we have found good 
convergence (low variation of independent runs) of simulations for this effective observable.

\section{Description of Simulations and Results}\label{sec:results}
Several sets of computations were performed in order to confirm the correctness of the algorithm and evaluate the
computational performance of the dual simulations as currently implemented. These are described in the 
subsections below; we first make some general comments.

Prior to computing the observable $\left< j \right>$, the conventional code was validated against published 
curves appearing in~\cite{LRSS1981}, which we selected because those simulations used the same heat kernel action 
employed in the present work.

Due to the computational cost of the vertex amplitude at higher spins, for the simulations described below
the series appearing in~(\ref{eqn:effectdef}) are cut off at a relatively low spin $j_{c}$; however, because
the cut-off is applied to both conventional (as a truncation of the character expansion) and dual forms, a direct
comparison can still be made, and posible errors in the (considerably more complex) dual code can be detected.
While low cut-off simulations can be quite accurate in the strong coupling regime, they begin to break down at the 
weaker coupling scales of interest. Further optimizations of the vertex amplitude underway by the author
should allow for higher spin cut-offs.

We now turn to a description of the numerical simulations performed on both sides of the duality that establish
the correctness (within statistical error) of the dual algorithm.  Error bars of $3\sigma$ are shown, where $\sigma$
is the standard deviation based on the average and variance of over the different simulation runs.

\subsection{Small lattice, small cut-off case}
Our first simulations were done on four-dimensional hypercubic lattice of $2^4$ vertices and a cutoff of $j_{c}=1$ (in units
of half-integer spin). The very small lattice size and low cut-off allowed relatively rapid simulations on both sides of the
duality in order to establish agreement.

For the conventional simulation, updating was done in ``sweeps'' in which each edge of the lattice was updated in sequence.
A total of $10^7$ sweeps were done at each value of $\beta$. For the dual simulations, a total of $2 \times 10^9$ moves 
were applied; these moves are described in the Appendix.
\begin{figure}
\includegraphics{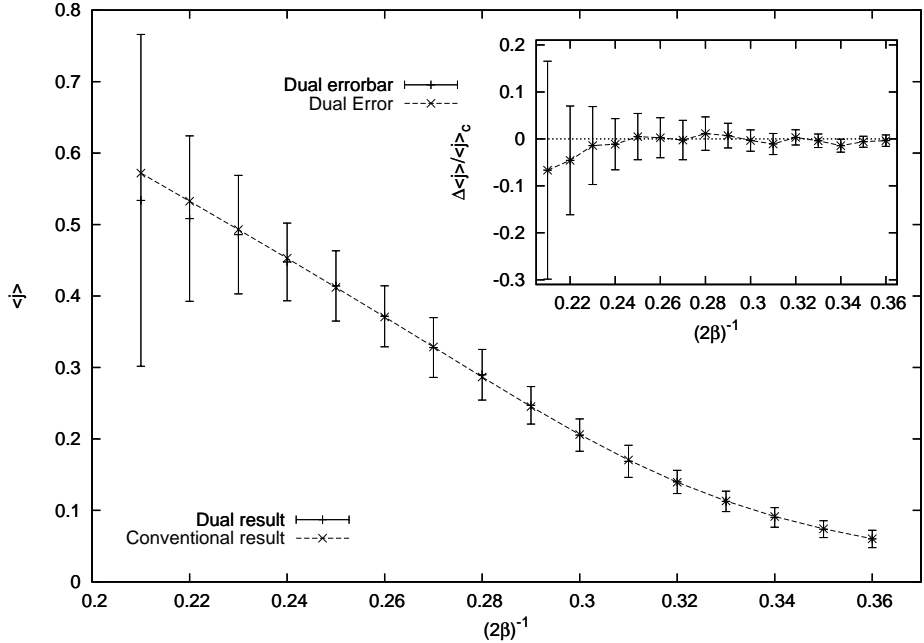}
\caption{Comparison of conventional with spin foam code for $2^4$ lattice at $j_{c}=1$ }\label{fig:compare_L2_J2}
\end{figure}
\subsection{Small lattice, cut-off sensitivity analysis}\label{sec:cutsense}
\begin{figure}
\includegraphics{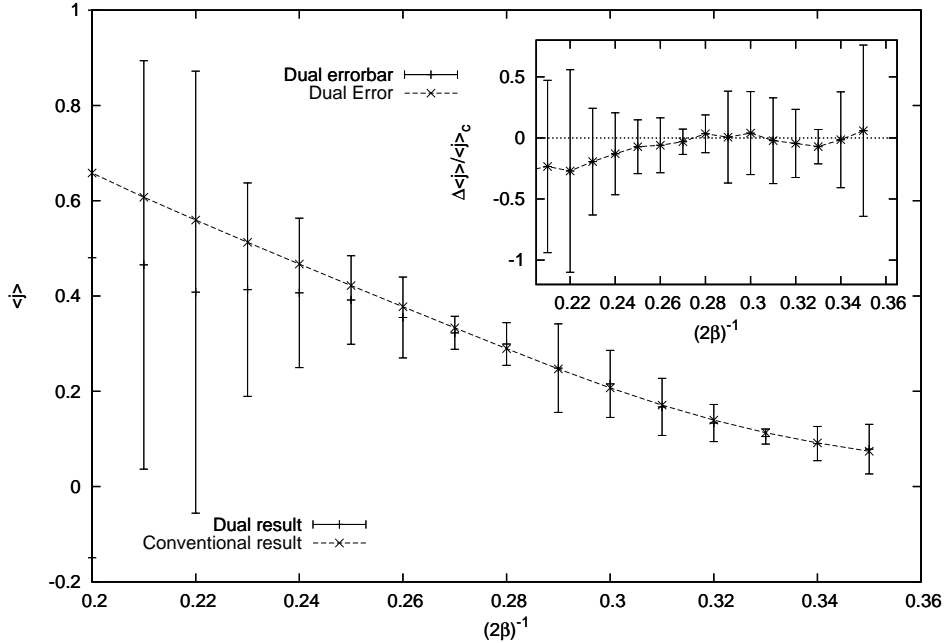}
\caption{Comparison of conventional with spin foam code for $2^4$ lattice at $j_{c}=\frac{3}{2}$ }\label{fig:compare_L2_J3}
\end{figure}

Further comparisons of dual and conventional simulations were made at increased spin cut-offs of $j_{c}=\frac{3}{2}$ 
and $j_{c}=2$. At these cut-offs, $1 \times 10^9$ samples were used in the dual simulations on the minimal ($2^4$) lattice. 
The results are shown in Figure~\ref{fig:compare_L2_J3} and again confirm the correctness of the dual algorithm
to within statistical error of the equivalent conventional algorithm.  Error bars at weak coupling are larger
at weak coupling, likely due to the smaller number of samples relative to the lower variance results obtained for $j_{c}=1$.
\subsection{Large lattice, small cut-off}\label{sec:largelat}
The dual algorithm was also investigated on larger lattice having dimensions $8^4$ at a spin cut-off of $j_{c}=1$.
At stronger coupling ($\beta = 1.8$ and lower) the dual simulations have high quality estimates with little relative
error.  However, for $\beta$ and greater than $1.8$ considerably larger errors begin to appear, rendering the results essentially
unusable. As expected, one clear source of error is in the expectation value of the sign, which at higher beta tends to smaller 
values. The error at $\beta=1.8$ on the $8^4$ lattice was 22\% and the sign expectation was of order $0.01$.  
For comparison, on the $2^4$ lattice the error at a comparable $\beta$ was 3\% and the sign expectation value was of the order $0.1$.
\section{Conclusions}
We have described the construction of a dual algorithm for $SU(2)$ Yang-Mills in $D=4$, based on a set of 
ergodic moves (suitable for use with a Metropolis or other algorithm) defined in the Appendix, and an efficient
form for the vertex amplitude, derived in~\cite{CC2009}.

Using a $2^4$ lattice as a simple test case, this algorithm was validated for a range of $\beta$ at spin
cut-offs of $j_c=1$ and $j_c=\frac{3}{2}$.  In this low-spin, small-lattice, intermediate coupling regime 
our algorithm exhibits statistical agreement with conventionally obtained results.
While this is clearly an impractically small size, it represents a milestone as the first dual simulation of its kind
in four space-time dimensions for a non-abelian gauge theory.

Towards strong coupling ($\beta$ of order unity), accurate results are possible even with 
the low cut-offs used (as large spins are penalized heavily by the amplitude in this regime).
However towards weak coupling an increased spin cut-off is required as configurations with larger spin
begin to dominate.  Because the computational cost of the dual vertex amplitude used increases with 
higher spins, raising the cut-off has the effect of slowing the rate at which samples are accumulated.
Thus, a way of more efficiently obtaining high accuracy results at weaker coupling would be to
speed up the vertex amplitude at high $j$, either through an improved algorithm, improved implementation,
or some combination of both. Note that, as demonstrated in the $D=3$ case~\cite{CCK}, different
choices of intertwiner bases and reductions of the spin network via recoupling moves lead to 
algorithms of different performance. Therefore exploring these different choices is a natural way
to seek out more efficient vertex amplitude algorithms. Intelligent caching of often used intermediate
calculations and the use of recursion relations that exist for $6j$ symbols are also being investigated
by the author.

A serious challenge for the new algorithm was encountered in increasing the lattice to $8^4$ in size. For
this lattice size the sign expectation value had a considerably more rapid fall-off with increased 
$\beta$.  As the overall sign of amplitude is the product of sign factors local vertex amplitudes, a 
decrease in the sign expectation value for any given $\beta$ can be generically expected as the 
lattice size becomes large, as was discussed in by Dass et al. in~\cite{DassShin}.
The rapid increased in error on data points for $\beta=1.8$ and greater exhibits has allowed
us to measure for the first time the extent of the sign problem for the four-dimensional case (within
our current approach).

A number of approaches are being pursued by the author to mitigate the problem of fluctuating signs. One general
approach would be to replace the $6j$ symbols that appear in the amplitude by their approximation (valid at large spins) in
the form of well-known the Ponzano-Regge asymptotic formula~\cite{PR68}.
While the Ponzano-Regge approximation itself contains fluctuating signs,  because the expressions are in closed
form they may useful as predictors of sign for the original $6j$ symbols and specifically to form a stationary
phase approximations, in which configurations of stationary phase are used rather than the full
ensemble~\cite{ConradyGluons}.  A possible problem with this approach is that the regime where asymptotic formulae for spins
may correspond to unrealistically large $\beta$.

Another approach to sign problems (that could work directly with the original vertex amplitude)
is to identify a sector of the original theory in which some form of cancellation occurs
in a controlled fashion (leading to a positive or at least better behaved sector for simulation). This has been
achieved for certain types of fermion models, under the rubric of \emph{meron cluster} and related methods; in some cases,
configurations with exactly equal weight and opposite sign can be identified and removed \emph{a priori}
from the simulation and the remaining ensemble sampled in an efficient manner~\cite{meroncluster}.

A general lesson for the sign problem is that while general methods that work in all instances don't
exist in tractable form~\cite{TRWI05}, by using details of how differently signed configuration
occur, it is possible to alleviate and in some casees completely remove the problem in a particular theory.
A final point to make is that a useful solution does not necessarily have to work for arbitrarily high $\beta$,
but rather mitigate the sign problem to the point of being practical for $\beta$ used to study length scales 
of particular interest.

In summary, the present work has demonstrated for the first time that dual simulations of Yang-Mills in
four space-time dimensions are feasible and provides a concrete, tested algorithm for carrying them out.
The main barriers to extending this approach to large lattice sizes and weaker coupling are the
emergence of a sign problem (the severity of which depends on lattice size) and the computational expense
of evaluating the vertex amplitude; some possible approaches to addressing these were identified.

\begin{acknowledgement*}
The author would like to thank Dan Christensen for valuable discussion influencing this work.
This work was made possible by the facilities of the Shared Hierarchical Academic Research Computing Network (SHARCNET)
\end{acknowledgement*}

\begin{appendices}
\section{Dual algorithm for $G=SU(2)$ in four dimensions}\label{sec:alg_append}

As in the conventional framework of LGT, the setting for the present algorithm is a 
hypercubic lattice with periodic boundary conditions such that the lattice has the 
topology of a 4-torus ($\mathbb{T}^4$).
The partition function for the theory can be expressed as a sum over dual amplitudes as
follows:
\begin{equation}\label{eqn:partfun}
Z_{YM} = \sum_{f \in F_{0}} \mathcal{A}(f) = \sum_{f \in F_{0}} \prod_{v \in V} A_{v}(\{j\}_v,\{i\}_v)
\prod_{e \in E} A_{e}(\{j\}_e,i_e) \prod_{p \in P} A_{p}(j_{p}),
\end{equation}
where $F_{0}$ is the set of lattice spin foam configurations (discussed further below). Each 
configuration labels a plaquette or edge by a quantum number. As can be seen from 
equation~(\ref{eqn:partfun}), the amplitude assigned to each spin foam factors into a product 
of contributions from local amplitudes associated to vertices, edges, and plaquettes of the lattice.
Recently, an explicit numerical algorithm was given in~\cite{CC2009} for each of these amplitudes in $D=4$.

For the practical evaluation of expectation values using numerical methods, it is often necessary to employ
Markov chain Monte Carlo techniques~\cite{Stirzaker} such as the Metropolis algorithm.
For such methods it is necessary to find a set of moves that are ergodic --- where any 
configuration can be transformed into any other by an appropriate sequence of the moves.  
For the present case, it can be non-trivial to find such moves due to the constrained nature 
of the dual configurations.

To describe the ergodic moves constructed for the present work, we must first identify the variables 
present at each vertex and edge, and the nature of the constraints they obey. Each plaquette carries a 
variable whose values are quantum numbers (we shall refer to them as ``spins'') that label the irreducible unitary 
representations of the group $G$.  Each edge carries an intertwiner: a $G$-invariant map 
between the irreducible representations assigned to the incident plaquettes.

In the present case of a four-dimensional hypercubic lattice, we have six incident plaquettes,
and hence the intertwiner must be an invariant map from a six-fold tensor product to the trivial representation.
Due to the self-duality of $SU(2)$ irreps, this is equivalent to an invariant map between a three-fold tensor 
product of irreps to a three-fold tensor product of irreps.
Intertwiners of this type form a three-dimensional vector space, and thus require three spin 
labels to resolve. The basis of intertwiners used can be represented diagrammatically as a
splitting of the six-valent vertex into a larger network with four three-valent vertices and
three ``internal'' edges. We note that there is some choice in how this splitting is carried out (equivalent to
changing the basis of intertwiners). As we are using the vertex amplitude described 
in~\cite{CC2009}, we shall adopt the corresponding choice of splitting, which is shown below in 
Figure~\ref{fig:splitting}. The three spin labels appear as $i_1$, $i_2$, and $i_3$ in 
Figure~\ref{fig:splitting}. The reader can readily check that these form an orthogonal 
basis by taking the inner product of two different labellings of the diagram by $i_{1},i_{2},i_{3}$. 
\begin{figure}[htb]
\includegraphics[scale=0.7]{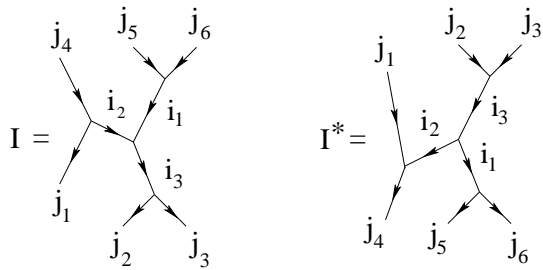}
\caption{Chosen splitting of the six-valent lattice edge and spin network vertex}\label{fig:splitting}
\end{figure}

On a four-dimensional hypercubic lattice, there are 24 plaquettes incident to each vertex and 8 edges. 
This leads to a vertex amplitude that depends on 24 spin labels coming from the plaquettes and $8 \times 3 = 24$
spin labels coming from the edges, in total a function of 48 spins.  The approach of fixing a basis and explicitly
carrying out a tensor contraction would be extraordinarily difficult to compute, with the cost scaling prohibitively with
the spin~\cite{CE2002}. However, as mentioned above, an algorithm that scales as $O(j^4)$ has recently 
been given in~\cite{CC2009}.

Having described the nature of the variables, it remains to define a set of ergodic moves.
To provide ergodicity, several distinct types of moves are applied.  As with the $D=3$ algorithm for $SU(2)$
constructed in~\cite{CCK}, these fall into three distinct classes --- homology moves, edge moves, and cube moves.
\begin{defn}[Homology move]
Select a plane of plaquettes that spans the entire lattice (and thus forms a closed surface due to the periodic 
boundary conditions). With equal probability, choose to decrease or increase a fundamental unit of spin and 
apply this change to every plaquette in the plane.
\end{defn}
Due to the periodic boundary conditions, these planes form closed surfaces that wrap the 4-torus. All admissible
deformations of these surfaces with the same globally wrapped topology are generated by the plaquette and edge moves.
The present algorithm uses a new approach to edge and plaquette moves that generally improves the acceptance
ratio compared to that of~\cite{CCK}. Recall that given any plaquette labelling, the allowed intertwiner
labels at each edge vary within a fixed range (non-empty if the plaquette labelling is admissible). Rather
than storing and applying moves to the intertwiner label itself, we store the offset of the intertwiner
relative to the minimum value in the admissible range. 

Specifically, given an edge splitting and labelling of incident plaquette spins $j_{1},\ldots,j_{6}$, the admissible 
intertwiner labellings $i_{1},i_{2},i_{3}$ are given by the intersection of the ranges defined by the
triangle inequalities. For example,  $ i_1 \in \left[| j_{5} - j_{6} |, j_{5} + j_{6} \right]$. For each edge, a 
relative intertwiner label $i^{r}$ for each of the three intertwiners is stored. In computing the amplitude of
the configuration, the intertwiner used is $i^{\text{min}} + i^{r}$ is used, where $i^{\text{min}}$ is the smallest
admissible intertwiner label.

We mention in passing that for certain special configurations, the amplitude may be zero despite the fact that all admissibility 
conditions are satisfied.  As was the case for the $D=3$ form of this algorithm~\cite{CCK}, we assume 
such configurations are sufficiently isolated from each other that they do not divide the configuration 
space into disconnected regions that can't be traversed by the moves.

\begin{defn}[Cube move]
A fundamental 3-cube of the lattice is selected at random. For each plaquette in the boundary of this cube, a
random choice is made to either increase or decrease the spin by a single fundamental unit of charge.
\end{defn}
Unlike the cube moves in~\cite{CCK}, no direct changes to the intertwiner labels is needed for compatibility.
The intertwiners labels are implicitly changed ``automatically'' by virtue of the admissible ranges
being modified by the change in plaquette label. Except when the spin is at the top of the admissible range 
and the plaquette moves reduce this range, a change in plaquette labels will implicitly give a new 
intertwiner labelling that is admissible. Another benefit of this approach is that it is actually easier to 
implement, as the algorithm doesn't have keep track of the different ways in which intertwiners sit inside
cubes (as was done in~\cite{CCK}), but simply which internal edges are connected to which plaquettes.

\begin{defn}[Edge move]
An edge $e$ and intertwiner label $i$ associated with that edge are randomly selected. The relative intertwiner
labelling $i^{r}$ is increased or decreased by two units of fundamental spin.
\end{defn}
We recall the change is by two units of spin due to the parity constraint; a single unit change in either direction
will always be inadmissible. 

In summary, the operation of the algorithm can be described as follows. An edge, cube, or homology move
type is selected at random and the proposed move is applied.  The contribution to the amplitude from those 
vertices, edges, and plaquettes effected by the move is computed before and after the move and the ratio of 
these local amplitudes is used to determine acceptance or rejection of the move by the standard Metropolis decision 
procedure.

\end{appendices}

\end{document}